 \documentclass[final,authoryear,5p,times,twocolumn]{elsarticle}
 
\usepackage{graphicx}

\usepackage{amssymb,amsmath}

\journal{Planetary and Space Science}

\begin{document}

\begin{frontmatter}



\title{Dust evolution with active galactic nucleus feedback in elliptical galaxies}


\author[label1]{Hiroyuki Hirashita}
\ead{hirashita@asiaa.sinica.edu.tw}
\author[label2]{Takaya Nozawa}
\address[label1]{Institute of Astronomy and Astrophysics,
Academia Sinica, P.O. Box 23-141, Taipei 10617, Taiwan}
\address[label2]{National Astronomical Observatory of Japan, Mitaka, Tokyo 181-8588, Japan}

\begin{abstract}
We have recently suggested that dust growth in the cold gas phase
dominates the dust abundance in elliptical galaxies while dust is
efficiently destroyed in the hot X-ray emitting plasma
(hot gas). In order to understand
the dust evolution in elliptical galaxies, we
construct a simple model that includes dust growth
in the cold gas and dust destruction in the hot gas.
We also take into account the effect of mass exchange between
these two gas components induced by active galactic nucleus
(AGN) feedback.
We survey reasonable ranges of the relevant parameters in
the model and find that AGN feedback
cycles actually produce a variety in cold gas mass and
dust-to-gas ratio.
By comparing with an observational sample of
nearby elliptical galaxies, we find that, although the dust-to-gas
ratio varies by an order of magnitude in our model,
the entire range of the observed dust-to-gas ratios is
difficult to be reproduced under a single
parameter set. Variation of the dust growth efficiency is
the most probable solution to explain the large variety in
dust-to-gas ratio of the observational sample.
Therefore, dust growth can play a central role in
creating the variation in dust-to-gas ratio through the AGN feedback
cycle and through the variation in dust growth efficiency.
\end{abstract}


\end{frontmatter}


\section{Introduction}
\label{sec:intro}

In the nearby Universe, elliptical galaxies are known to have less gas, dust,
and ongoing star formation activity than spiral galaxies. Yet, they
still have some amount of interstellar gas in the form of hot X-ray-emitting
halo gas \citep[e.g.,][]{Osullivan:2001aa} and cold gas
\citep[e.g.,][]{Wiklind:1995aa}.
Moreover, dust is detected for a significant fraction of elliptical galaxies by
optical extinction \citep[e.g.,][]{Goudfrooij:1994aa,van-Dokkum:1995aa,Ferrari:1999aa,Tran:2001aa}
or far-infrared (FIR) emission \citep[e.g.,][]{Knapp:1989aa,Smith:2012aa,di-Serego-Alighieri:2013aa}.
Dust mass is estimated from the reddening in the optical or from FIR
dust emission, and  ranges from $\sim 10^4$ to $\sim 10^7~M_\odot$
\citep[e.g.,][]{Goudfrooij:1994aa,Leeuw:2004aa}. Since
the existence of dust could affect the cooling and chemical processes
\citep{Dwek:1987aa,Fabian:1994aa,Voit:1995aa,Seok:2015aa,Hirashita:2015aa},
the understanding of the origin and evolution of
dust in elliptical galaxies is important in clarifying their evolution.

Because the stellar population is dominated by old stars whose ages are
comparable to the cosmic age, the dust in elliptical galaxies is predominantly supplied by
asymptotic giant branch (AGB) stars rather than by supernovae. However,
dust destruction by sputtering in the X-ray emitting plasma
is so efficient that the observed dust mass cannot be explained
by the balance between the supply
from AGB stars and the destruction \citep[e.g.,][]{Patil:2007aa}.
Thus, some authors argue that the dust existing in elliptical galaxies
is possibly injected from outside via the merging or accretion of
external galaxies \citep{Forbes:1991aa,Temi:2004aa,Fujita:2013aa}.
The lack of correlation between dust FIR luminosity and stellar luminosity
is also taken as evidence of external origin of dust
\citep{Temi:2007aa}; however, this argument may not hold if dust is
processed by mechanisms unrelated to stars.

Recently, \citet{Hirashita:2015aa} have proposed that the existence
of dust in elliptical galaxies can be explained by dust growth by
the accretion of gas phase metals in the cold interstellar medium (ISM).
They also suggest that the presence of dust growth also explains
the extinction curves observed in elliptical
galaxies. The dust-to-gas ratio could become as high as
$\gtrsim 10^{-3}$ by accretion, explaining the high dust abundance
in some elliptical galaxies. However, dust growth in the
cold gas has not been considered as a major source of dust in the
context of dust evolution in elliptical galaxies.
Dust growth has already been noted as a major mechanism of
dust mass increase in a wide context of galaxy evolution
\citep[e.g.,][]{Dwek:1998aa,Hirashita:1999aa,Inoue:2003aa,Zhukovska:2008aa,
Valiante:2011aa,Mattsson:2012aa,Mancini:2015aa,Popping:2016aa,Hou:2016aa,Zhukovska:2016aa},
and some experimental studies have also shown that dust grains
could grow by accreting the gas-phase metals
(\citealt{Rouille:2014aa}; but see \citealt{Ferrara:2016aa}).
Thus, constructing a
dust evolution model that includes this dust growth mechanism
would contribute to the understanding of dust evolution in elliptical
galaxies.

The overall evolution of dust should be considered in relation to
the gas evolution, especially because gas and dust are usually
dynamically coupled on galactic scales. Recent studies have
proposed that the
gas cooling in galaxies is strongly regulated by the energy input
from active galactic nuclei (AGNs). The energy input from AGN winds
\citep{Silk:1998aa,Fabian:1999aa,King:2005aa} or AGN jets
\citep{Wagner:2011aa,Mukherjee:2016aa} prevents cooling flows
from occurring and/or makes cold gas evaporate
\citep[see also][]{Ciotti:2001aa,Fabian:2012aa}.
These kinds of energy input are called AGN feedback, and are
considered to play an important role in galaxy formation
and evolution \citep{Croton:2006aa,Booth:2009aa}

\citet{Temi:2007aa} showed the existence of a dust emission
component extended over 5--10~kpc in elliptical galaxies using
the \textit{Spitzer} 70 $\mu$m band data. They suggest that
this cold dust component was originally
contained in the cold gas, which was then heated by AGN feedback and
eventually mixed with the hot gas. In their scenario, the heated gas
is transported into the hot gas
by buoyant force, and the time-scale of the transport is around $10^7$ yr.
\citet{Kaneda:2011aa} also showed in an elliptical galaxy (NGC 4125)
a dust emission component whose
extension is similar to the distribution of the hot X-ray emitting halo.
The multi-phase structures and irregular morphologies of
X-ray emitting hot plasmas could be explained by bubbles
created by AGN feedback \citep{Buote:2003aa}.

A part of the hot gas may cool down to reform the cold gas, which
could contribute to the fueling of AGN \citep{Werner:2014aa}.
AGN feedback may compress the surrounding gas and enhance
gas cooling locally \citep{Valentini:2015aa}. Because dust grains
can grow in the cold gas as mentioned above, the cold gas
injected into the hot gas by AGN feedback would supply the dust
to the hot gas. This dust supply could be dominant
over the dust production by AGB stars. If so, formation of cold gas and
the subsequent AGN fueling leading to AGN feedback play a
dominant role in the dust evolution in elliptical galaxies.

If the dust evolution is affected by episodic AGN activities,
which appear as a result of a cycle of gas ejection and cooling,
the statistical properties of the dust abundances in elliptical galaxies
are determined by the time-scales of dust processing relative to the
period of an AGN cycle. \citet{Lauer:2005aa} inferred the period of
an AGN cycle in early-type galaxies based on the dust lifetime against sputtering in the hot
gas and the fraction of dust detection for a sample, obtaining a period of $\sim 10^8$ yr.
The dust contained in the cold gas may be injected into the hot
gas in an episodic way associated with the AGN cycle, if AGN feedback
efficiently heats the cold gas \citep{Mathews:2013aa}.

In this paper, we make a theoretical model of dust evolution in AGN feedback
cycles by including important physical processes such as
the mass exchange between the hot and cold gas components
and the dust evolution in those gases. For dust processing, we
consider dust destruction by sputtering in the hot gas and dust
growth by the accretion of gas-phase metals in the cold gas.
This modeling enables us to examine how AGN feedback cycles
affect the dust abundances in elliptical galaxies.
We can also examine the effect of dust growth on the dust abundance in
elliptical galaxies for the first time.

The paper is organized as follows: we formulate the model
of dust evolution in an elliptical galaxy in Section~\ref{sec:model}.
We show the results in Section~\ref{sec:result}. We discuss
the model predictions, and compare them with
observational data in Section \ref{sec:discussion}.
Finally we conclude in Section \ref{sec:conclusion}.

\section{Model}\label{sec:model}

We construct a model that describes the dust evolution in
an elliptical galaxy. For dust sources, in addition to AGB
stars considered in previous studies \citep{Mathews:2003aa,Patil:2007aa},
we also consider dust supply from the cold gas where
dust grows by the accretion of gas-phase metals. The dust
in the cold gas is injected into
the hot gas by AGN feedback together with the cold gas.
The dust injected into the hot gas is destroyed by sputtering.
The hot component ($\sim 10^7$ K) is the gas whose temperature
is comparable to the virial temperature determined by the
global gravitational potential of the elliptical galaxy while
the cold component is cold and dense enough to host dust growth
\citep[$\lesssim 100$ K;][]{Hirashita:2015aa}.
To make the model as simple as possible, we treat each gas
component as a single zone and consider the mass exchange between the
two components. Dust growth and destruction are treated consistently
with the evolution of each gas component as explained below.

\subsection{Basic equations}\label{subsec:basic}

We consider the mass exchange between the hot and cold phases.
The evolution of the cold gas mass $M_\mathrm{g,c}$ as a function
of time $t$ is written as
\begin{eqnarray}
\frac{dM_\mathrm{g,c}}{dt}=\dot{M}_\mathrm{in}-\dot{M}_\mathrm{ret},
\label{eq:Mgc}
\end{eqnarray}
where $\dot{M}_\mathrm{in}$ is the infall rate of the cooled
gas and $\dot{M}_\mathrm{ret}$ is the gas return rate from
the cold to the hot phase by AGN feedback. These two terms are
modeled below.

The hot gas is treated as a constant reservoir of gas  for simplicity;
that is,
we assume that the mass of hot gas, $M_\mathrm{g,h}$ is constant
($dM_\mathrm{g,h}/dt=0$). This treatment
neglects the complication arising from the possibility 
that there could be a supply/loss of hot gas
from/to outside. Because the hot gas basically acts as an efficient
destroyer of dust regardless of the choice of $M_\mathrm{g,h}$,
the value of $M_\mathrm{g,h}$ has a minor
influence on the results compared with other parameters.
However, we should note that the dust-to-gas ratio in the hot gas
is directly affected by $M_\mathrm{g,h}$ since it directly enters
the dust-to-gas ratio in the hot gas. We discuss the value of
$M_\mathrm{g,h}$ again in Section \ref{subsec:param}.

The time evolution of the dust mass in the cold phase, $M_\mathrm{d,c}$,
is written as
\begin{eqnarray}
\frac{dM_\mathrm{d,c}}{dt}=\mathcal{D}_\mathrm{h}\dot{M}_\mathrm{in}-
\mathcal{D}_\mathrm{c}\dot{M}_\mathrm{ret}+
\frac{M_\mathrm{d,c}}{\tau_\mathrm{grow}},\label{eq:Mdc}
\end{eqnarray}
where $\mathcal{D}_\mathrm{h}\equiv M_\mathrm{d,h}/M_\mathrm{g,h}$
and $\mathcal{D}_\mathrm{c}\equiv M_\mathrm{d,c}/M_\mathrm{g,c}$ are
the dust-to-gas ratios in the hot and cold phases, respectively
($M_\mathrm{d,h}$ and $M_\mathrm{d,c}$ are the dust mass in the
hot and cold gas, respectively),
and $\tau_\mathrm{grow}$ is the time-scale of dust growth by the
accretion of gas-phase metals in the cold gas.
{The dust growth timescale $\tau_\mathrm{grow}$ is given
in Section \ref{subsec:param} (equation \ref{eq:taugrow}), and is treated as a function of
$\mathcal{D}_\mathrm{c}$.}
The time evolution of the dust mass in the hot phase, $M_\mathrm{h,c}$,
on the other hand, is written as
\begin{eqnarray}
\frac{dM_\mathrm{d,h}}{dt}=-\mathcal{D}_\mathrm{h}\dot{M}_\mathrm{in}
+\mathcal{D}_\mathrm{c}\dot{M}_\mathrm{ret}-
\frac{M_\mathrm{d,h}}{\tau_\mathrm{sput}}+\mathcal{D}_\mathrm{AGB}\alpha M_*,
\label{eq:Mdh}
\end{eqnarray}
where $\tau_\mathrm{sput}$ is the dust destruction time-scale by
sputtering in the hot gas, $\mathcal{D}_\mathrm{AGB}$ is the dust-to-gas
ratio in AGB star winds, $\alpha$ is the mass loss rates of AGB stars
per stellar mass, and $M_*$ is the total stellar mass
($\alpha M_*$ is the total mass loss rate of AGB stars).
Using $M_\mathrm{d,c}=\mathcal{D}_\mathrm{c}M_\mathrm{g,c}$,
$M_\mathrm{d,h}=\mathcal{D}_\mathrm{h}M_\mathrm{g,h}$, and
equation (\ref{eq:Mgc}), we rewite equations (\ref{eq:Mdc}) and
(\ref{eq:Mdh}) as the following equations for the dust-to-gas ratios:
\begin{eqnarray}
\frac{d\mathcal{D}_\mathrm{c}}{dt}=
\frac{\mathcal{D}_\mathrm{c}}{\tau_\mathrm{grow}}+
(\mathcal{D}_\mathrm{h}-\mathcal{D}_\mathrm{c})
\frac{\dot{M}_\mathrm{in}}{M_\mathrm{g,c}},\label{eq:Dc}
\end{eqnarray}
\begin{eqnarray}
\frac{d\mathcal{D}_\mathrm{h}}{dt}=
-\mathcal{D}_\mathrm{h}\frac{\dot{M}_\mathrm{in}}{M_\mathrm{g,h}}
-\frac{\mathcal{D}_\mathrm{h}}{\tau_\mathrm{sput}}
+\mathcal{D}_\mathrm{c}\frac{\dot{M}_\mathrm{ret}}{M_\mathrm{g,h}}
+\mathcal{D}_\mathrm{AGB}\frac{\alpha M_*}{M_\mathrm{g,h}}.\label{eq:Dh}
\end{eqnarray}

We solve equations (\ref{eq:Mgc}), (\ref{eq:Dc}) and (\ref{eq:Dh}).
Below we formulate some undetermined terms and estimate a
reasonable range for each parameter (Table \ref{tab:param}).

\begin{table}
\centering
\begin{minipage}{80mm}
\caption{Parameters in the model.}
\label{tab:param}
\begin{center}
\begin{tabular}{@{}lccc} \hline
Parameter & Unit & Fiducial$^a$ & Range$^b$
\\ \hline
$\dot{M}_\mathrm{in,0}$ & $M_\odot$ yr$^{-1}$ & 3 & 1--10\\
$\tau_\mathrm{tr}$ & yr & $10^{7}$ & (0.3--3)${}\times 10^7$\\
$\tau_\mathrm{AGN}$ & yr & $10^{8}$ & fixed\\
$f_\mathrm{ret}$ & --- & 0.3 & 0.1--0.5 \\
$\tau_\mathrm{acc}$ & yr & $10^7$ & $10^6$--$10^8$ \\
$\tau_\mathrm{sput}$ & yr & $10^7$ & $10^6$--$10^8$ \\
$Z$ & --- & 0.02 & fixed \\
$\mathcal{D}_\mathrm{AGB}$ & --- & 0.01 & fixed \\
$\alpha$ & yr$^{-1}$ & $1.6\times 10^{-12}$ & fixed \\
$M_*$ & $M_\odot$ & $10^{10.5}$ & $10^{10}$--$10^{11}$ \\
\hline
\end{tabular}
\end{center}
$^a$Fiducial value.\\
$^b$Range of values investigated unless the parameter value is fixed.
\end{minipage}
\end{table}

\subsection{Mass exchange between the phases}\label{subsec:exchange}

The mass exchange between the cold and hot phases is described
by $\dot{M}_\mathrm{in}$ (inflow of cooled hot gas to the cold gas)
and $\dot{M}_\mathrm{ret}$ (return of cold gas to the hot phase by
AGN feedback). For simplicity, we assume that AGN feedback
occurs periodically with a period of $\tau_\mathrm{AGN}$.
During an episode of AGN feedback, the cold gas is transported
to the hot gas, and this transportation time-scale is parameterized by
$\tau_\mathrm{tr}$. This transport (or the return of cold gas to the
hot gas) lasts for a fraction of $\tau_\mathrm{AGN}$, and
this fraction is parameterized by $f_\mathrm{ret}$. That is,
the duration of an AGN feedback episode is $f_\mathrm{ret}\tau_\mathrm{AGN}$.
We assume that
the outflow ($\dot{M}_\mathrm{ret}$) of the cold gas is on during the AGN feedback as
\begin{eqnarray}
\dot{M}_\mathrm{ret}(t)=\left\{
\begin{array}{ll}
{M_\mathrm{g,c}}/{\tau_\mathrm{tr}} &
\mbox{if $0\leq t/\tau_\mathrm{AGN}-[t/\tau_\mathrm{AGN}]\leq f_\mathrm{ret}$,}\\
0 & \mbox{otherwise,}
\end{array}
\right.\label{eq:Mret}
\end{eqnarray}
where $[x]$ is the floor function, which indicates the
largest integer that satisfies $[x]\leq x$.
{This equation indicates that the outflow is periodically turned on
every $\tau_\mathrm{AGN}$ with a duration of
$f_\mathrm{ret}\tau_\mathrm{AGN}$.}

The inflow of cooled hot gas is assumed to occur only when there
is no outflow activity (i.e., when $M_\mathrm{ret}=0$):
\begin{eqnarray}
\dot{M}_\mathrm{in}(t)=\left\{
\begin{array}{ll}
0 &
\mbox{if $0\leq t/\tau_\mathrm{AGN}-[t/\tau_\mathrm{AGN}]\leq f_\mathrm{ret}$,}\\
\dot{M}_\mathrm{in,0} & \mbox{otherwise,}
\end{array}
\right.
\end{eqnarray}
where we treat $\dot{M}_\mathrm{in,0}$ as a constant free parameter.

Based on the inferred dust lifetime and dynamical time in the
center of an elliptical galaxy and the rate of dust detection for
elliptical galaxies, \citet{Lauer:2005aa} suggested that the cycle of  dust
appearance and disappearance is $\sim 10^8$ yr.
We can regard this time as $\tau_\mathrm{AGN}$ if that dust
cycle is induced by the AGN feedback cycle. Thus, we adopt
$\tau_\mathrm{AGN}=10^8$ yr. Since the time evolution of gas and
dust is determined by the time-scale relative to $\tau_\mathrm{AGN}$,
we fix $\tau_\mathrm{AGN}$ and change other time-scales as
described below.
The maximum cold gas mass is roughly estimated as
$\dot{M}_\mathrm{in,0}\tau_\mathrm{AGN}(1-f_\mathrm{ret})$,
which should be comparable to the observed gas mass
$\sim\mbox{a few}\times 10^8~M_\odot$
\citep[e.g.,][]{Wiklind:1995aa}. This means that
$\dot{M}_\mathrm{in,0}\sim\mbox{a few}\times (1-f_\mathrm{ret})^{-1}~M_\odot$ yr$^{-1}$.
Thus, we investigate a range of $\sim 1$--10 $M_\odot$ yr$^{-1}$.
Since cold gas is detected in a significant fraction of elliptical galaxies,
$f_\mathrm{ret}$ cannot be very near to 1. Thus, we focus on a range of
$0<f_\mathrm{ret}\leq 0.5$.

For $\tau_\mathrm{tr}$, it may be reasonable to adopt the dynamical
time. As mentioned in the Introduction, \citet{Temi:2007aa} suggested dust transport by buoyancy
acting on the heated cold gas in order to explain the origin of
extended dust FIR emission component.
They estimated the time-scale of transport as $\sim 10^7$ yr
based on the dynamical time-scale. Thus, we adopt
$\tau_\mathrm{tr}\sim 10^7$ yr, and also examine an order-of-magnitude
range for $\tau_\mathrm{tr}$.

\subsection{Parameters for dust evolution}\label{subsec:param}

We have introduced two time-scale parameters that govern
the dust evolution: dust growth time-scale ($\tau_\mathrm{grow}$)
and dust destruction (sputtering) time-scale ($\tau_\mathrm{sput}$).

We adopt the following estimate of the time-scale of grain
growth \citep{Asano:2013ab}:
\begin{eqnarray}
\tau_\mathrm{acc} & = & 2\times 10^6\left(\frac{a}{0.1~\mu\mathrm{m}}\right)
\left(\frac{n_\mathrm{H,c}}{10^3~\mathrm{cm}^{-3}}\right)^{-1}
\left(\frac{T_\mathrm{gas}}{50~\mathrm{K}}\right)^{-1/2}\nonumber\\
& & \times\left(\frac{Z}{0.02}\right)^{-1}~\mathrm{yr},
\end{eqnarray}
where $a$ is the grain radius, $n_\mathrm{H,c}$ is the hydrogen number
density in the cold gas, $T_\mathrm{gas}$ is the gas temperature, and
$Z$ is the metallicity
\citep[e.g.,][]{Inoue:2011aa,Asano:2013ab}.
{Because of some poorly constrained parameters such as
$a$ and $n_\mathrm{H,c}$, we treat $\tau_\mathrm{acc}$ as a given
constant parameter for simplicity, but investigate a wide range
in $\tau_\mathrm{acc}$. Moreover, since}
accretion is efficient only in the dense
medium, it depends on the mass fraction of dense gas in the cold gas,
which is denoted as $f_\mathrm{dense}$. The dust growth time-scale is
effectively estimated as $\tau_\mathrm{acc}/f_\mathrm{dense}$.
Based on the above estimate, we
examine $\tau_\mathrm{acc}=10^6$--$10^8$ yr
($10^7$ yr for the fiducial case), allowing for a small value for
$f_\mathrm{dense}$ or variation in $a$ and $n_\mathrm{H,c}$
\citep{Kuo:2013aa,Schneider:2016aa,Aoyama:2017aa}.
In reality, the accretion time-scale is also affected by
grain size distribution
\citep{Kuo:2012aa}; {thus, the above $a$ represents the appropriate
average of the grain size.\footnote{More precisely, $a$ is the
ratio of the mean $a^3$ to the mean $a^2$ \citep{Hirashita:2011aa}.
As shown in \citet{Hirashita:2011aa}, the time variation of
$a$ could be effectively incorporated in the definition of $\tau_\mathrm{acc}$
so that $\tau_\mathrm{acc}$ could still be treated as constant.}
Considering the uncertainties in grain size distribution,
$\tau_\mathrm{acc}$ may be out of the range above ($10^6$--$10^8$ yr).
If $\tau_\mathrm{acc}\lesssim 10^6$ yr, the
grain growth is so efficient that the dust-to-gas ratio in the cold
gas stays at a
maximum value given below ($\mathcal{D}_\mathrm{sat}$) for most of the
time regardless of the value of $\tau_\mathrm{acc}(\lesssim 10^6~\mathrm{yr})$.
If $\tau_\mathrm{acc}\gtrsim 10^8$ yr, in contrast, dust growth is
not efficient enough to raise the dust-to-gas ratio above
$10^{-4}$. Therefore, the range of $\tau_\mathrm{acc}=10^6$--$10^8$ yr
covers all the resulting behavior of interest for $\mathcal{D}_\mathrm{c}$,
i.e., covers all the interesting range of $\mathcal{D}_\mathrm{c}$
as we show below.}

{Using a given value of $\tau_\mathrm{acc}$, the dust growth time-scale is
given by}
\begin{eqnarray}
\tau_\mathrm{grow}=
\frac{\tau_\mathrm{acc}}{1-\mathcal{D}_\mathrm{c}/\mathcal{D}_\mathrm{sat}},
\label{eq:taugrow}
\end{eqnarray}
where the denominator on the right-hand side means that
accretion is saturated as $\mathcal{D}_\mathrm{c}$ approaches
$\mathcal{D}_\mathrm{sat}$, which is the abundance of the metals
available for dust formation. We assume $\mathcal{D}_\mathrm{sat}=0.01$
in this paper,
since we are considering solar metallicity environment
(i.e., $\mathcal{D}_\mathrm{sat}\sim Z_\odot$).
{The abundance of available metals for dust formation may be
correlated with the stellar mass because of the stellar mass--metallicity
relation \citep{Gallazzi:2005aa}. However, this relation has a large scatter,
and it is also suggested that the velocity dispersion (or the virial mass) is
more tightly related to the stellar metallicity \citep{Smith:2009aa}.
Moreover, it is not clear how tightly the stellar metallicity is related to
the gas metallicity because of accretion of external gas \citep{Su:2013aa}.
Thus, we simply adopt a constant $\mathcal{D}_\mathrm{sat}$,
but we also note that increasing/decreasing $\mathcal{D}_\mathrm{sat}$
has broadly the same effect of short/long $\tau_\mathrm{acc}$.}

The time-scale of dust destruction by sputtering is estimated as
\citep{Tsai:1995aa,Nozawa:2006aa,Hirashita:2015aa}
\begin{eqnarray}
\tau_\mathrm{sput}=7.1\times 10^6
\left(\frac{a}{0.1~\mu\mathrm{m}}\right)
\left(\frac{n_\mathrm{H,h}}{10^{-2}~\mathrm{cm}^{-3}}\right)^{-1}~\mathrm{yr},
\end{eqnarray}
where $n_\mathrm{H,h}$ is the number density of hydrogen nuclei
in the hot gas. 
{As we did for $\tau_\mathrm{acc}$ above, we treat
$\tau_\mathrm{sput}$ as a given constant parameter.}
Based on this estimation, we adopt
$\tau_\mathrm{sput}=10^7$ yr for the fiducial case
and examine a range of $10^6$--$10^8$ yr considering the variety in
$a$ and $n_\mathrm{H,h}$.
{Note that fixing $\tau_\mathrm{acc}$ and $\tau_\mathrm{sput}$
implicitly assumes a fixed grain size (or grain size distribution).
In reality, the grain size distribution itself is
determined as a result of the cycle of dust destruction and growth.
Therefore, a consistent treatment of those time-scales and
the grain size distribution would be necessary.
\citet{Hirashita:2015aa} considered the evolution of grain size
distribution by dust destruction and subsequent dust growth and showed that
the resulting grain size distribution is
consistent with the optical extinction curves observed by
\citet{Patil:2007aa}, although they did not consider a full cycle
of AGN feedback. At the same time,
they also showed that the resulting grain size distribution depends on
the size distribution of dust grains produced by
AGB stars. To avoid such a complication, we treat
$\tau_\mathrm{acc}$ and $\tau_\mathrm{sput}$ as given constant parameters
and concentrate on the total dust mass.
The grain size distribution realized after AGN feedback
cycles will be investigated in the future work.}

We adopt $\mathcal{D}_\mathrm{AGB}=0.01$,
$\alpha =1.6\times 10^{-12}$ yr$^{-1}$, and
$M_*=10^{10}$--$10^{11}~M_\odot$
\citep{Tsai:1995aa}.
However, the precise values of these quantities do not affect
the dust abundance in the cold gas as long as the dust formed in AGB stars
is efficiently destroyed by the hot gas.
We fix $\mathcal{D}_\mathrm{AGB}$ and $\alpha$,
and only focus on the variation of $M_*$.

As explained in Section \ref{subsec:basic}, we fix the
hot gas mass $M_\mathrm{g,h}$ for simplicity.
The hot gas mass
in the central a few kpc, which could directly exchange the
gas mass with the central cold component, is estimated as
$10^8$--$10^9~M_\odot$ \citep{Tsai:1995aa}.
Thus, we adopt $M_\mathrm{g,h}=3\times 10^8~M_\odot$.
The dust-to-gas ratio in the hot gas ($\mathcal{D}_\mathrm{h}$)
is inversely proportional
to $M_\mathrm{g,h}$ by definition; however, because dust in
the hot gas is efficiently destroyed by sputtering, as shown below,
$\mathcal{D}_\mathrm{h}\ll\mathcal{D}_\mathrm{c}$ is
satisfied when the hot gas is actively accreted on to the cold gas,
except for the case where the sputtering time-scale is as long as
the AGN feedback cycle ($\tau_\mathrm{sput}\gtrsim\tau_\mathrm{AGN}$).
Therefore,
the inflow is practically a ``dust-free'' gas for the cold gas,
so that the value of $\mathcal{D}_\mathrm{h}$ does not influence
$\mathcal{D}_\mathrm{c}$. This means that
the choice of $M_\mathrm{g,h}$, which could affect $\mathcal{D}_\mathrm{h}$,
does not have a significant influence on $\mathcal{D}_\mathrm{c}$.

We assume the following initial conditions:
$\mathcal{D}_\mathrm{c}=\mathcal{D}_\mathrm{h}=10^{-6}$, and
$M_\mathrm{g,c}=10^7~M_\odot$. As seen later, the initial
condition is not important, since the system
``forgets'' the initial condition in $\sim \tau_\mathrm{AGN}$.

\section{Results}\label{sec:result}

\subsection{Fiducial case}

\begin{figure}
\includegraphics[width=0.45\textwidth]{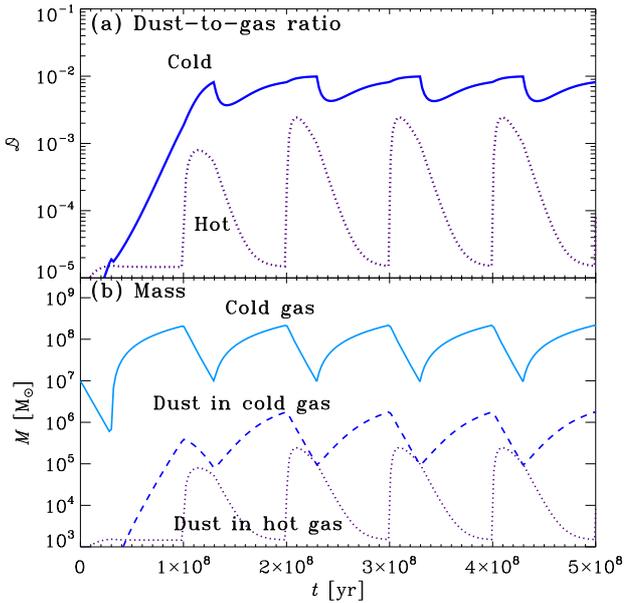}
\caption{Time evolutions of relevant quantities.
(a) Dust-to-gas ratio in the cold ($\mathcal{D}_\mathrm{c}$)
and hot gas ($\mathcal{D}_\mathrm{h}$)
(solid and dotted lines, respectively).
(b) Cold gas mass (${M}_\mathrm{g,c}$) (solid line),
and dust masses
in the cold gas ($M_\mathrm{d,c}$) (dashed line) and in the hot gas
($M_\mathrm{h,c}$) (dotted line).
\label{fig:std}}
\end{figure}

First of all, we show the time evolution of relevant
quantities under the fiducial parameter values listed in
Table \ref{tab:param}. In Fig.~\ref{fig:std},
we show the evolutions of the dust-to-gas ratios in
the hot and cold gas, the cold gas mass, and the
dust masses in the cold and hot gas. Here we describe
how each of these quantities evolves as a function of
time.

All the quantities have almost periodic behaviors
with a period given by $\tau_\mathrm{AGN}$.
In association with the onset of AGN feedback, the
dust grown in the cold gas is injected into the
hot gas, so that the dust-to-gas ratio in the hot
gas increases. However, the dust supplied to the hot gas
is quickly destroyed by sputtering,
{and the dust-to-gas
in the hot gas tends to converge to the value determined
by the balance between dust supply from stars and dust
destruction by sputtering}.
After AGN feedback stops, the
cold gas increases its mass by gas infall.
In this phase, because the dust-to-gas ratio
in the infalling gas, which is equal to the dust-to-gas ratio
in the hot gas, is smaller than that in the cold gas,
the dust-to-gas ratio in the cold gas decreases
as a result of \textit{dilution}.

The above general behavior does not change by the choice of
parameter values. Thus, for the purpose of simplification,
we hereafter concentrate on
the dust-to-gas ratios in the cold and hot gas
($\mathcal{D}_\mathrm{c}$ and $\mathcal{D}_\mathrm{h}$),
and the cold gas mass ($M_\mathrm{g,c}$), since the
dust masses in the hot and cold gas just trace the
evolution of the corresponding dust-to-gas ratios.
Below we show the dependence on each parameter listed
in Table \ref{tab:param}.

\subsection{Mass inflow rate}\label{subsec:Min}

\begin{figure}
\includegraphics[width=0.45\textwidth]{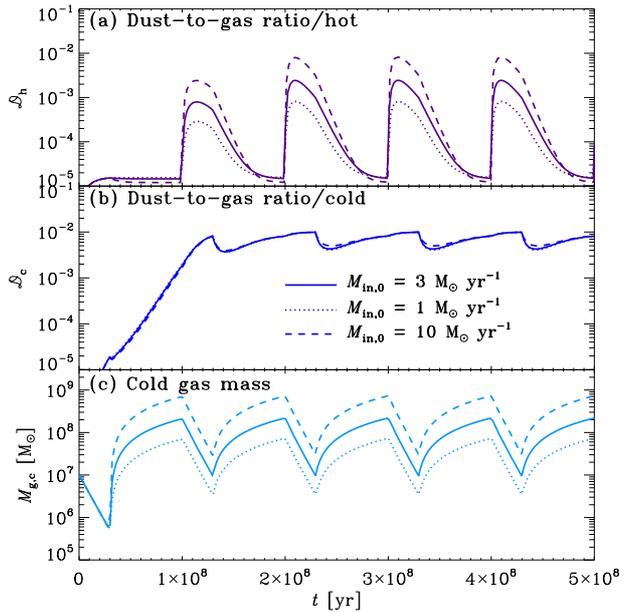}
\caption{Time evolutions of relevant quantities.
(a) Dust-to-gas ratio in the hot gas ($\mathcal{D}_\mathrm{h}$).
(b) Dust-to-gas ratio in the cold gas ($\mathcal{D}_\mathrm{c}$).
(c) Cold gas mass (${M}_\mathrm{g,c}$).
The solid, dotted, and dashed lines show the results for
$\dot{M}_\mathrm{in,0}=3$ (fiducial), 1, and 10 $M_\odot$,
respectively, in all the panels.
\label{fig:Min}}
\end{figure}

We consider the mass accretion onto the cold gas when
AGN feedback is off. The mass inflow rate is regulated by
the parameter $\dot{M}_\mathrm{in,0}$. In Fig.\ \ref{fig:Min},
we show the time evolution of $\mathcal{D}_\mathrm{c}$,
$\mathcal{D}_\mathrm{h}$, and $M_\mathrm{g,c}$ for
$\dot{M}_\mathrm{in,0}=3$ (fiducial), 1, and 10 $M_\odot$ yr$^{-1}$.
Below we describe how the results depend on 
$\dot{M}_\mathrm{in,0}$.

The dust-to-gas ratio in the hot gas has a larger amplitude
for a larger $\dot{M}_\mathrm{in,0}$. This is because
more dust is accumulated in the cold gas. When this dusty cold gas
is injected into the hot gas by AGN feedback, the hot gas is
enriched with dust.
The dust injected into the hot gas is readily destroyed by
sputtering; thus, the hot gas is dust-poor for most of the time.
The dust-to-gas ratio in the hot gas drops slightly more
in the case of a higher $M_\mathrm{in,0}$ because the inflow
transports the dust from the hot gas to the cold gas in our
model. However, the lowest level of the dust-to-gas ratio in the
hot gas is not very sensitive to $M_\mathrm{in,0}$ since it is
determined by the balance between dust supply from AGB stars
and dust destruction by sputtering.
In terms of the dust-to-gas ratio in the cold gas,
the periodic decrease is due to the inflow of dust-poor gas.
However, since dust growth is efficient enough, the
original high dust-to-gas ratio is recovered as soon as
the cold gas attains enough mass by the inflow. As expected,
the cold gas mass is larger if $\dot{M}_\mathrm{in,0}$
is larger, but the logarithmic amplitude of the cold
gas mass does not depend on $\dot{M}_\mathrm{in,0}$, since
we fix the time-scales of AGN feedback and mass transport.

\subsection{Mass transport time-scale in AGN feedback}

\begin{figure}
\includegraphics[width=0.45\textwidth]{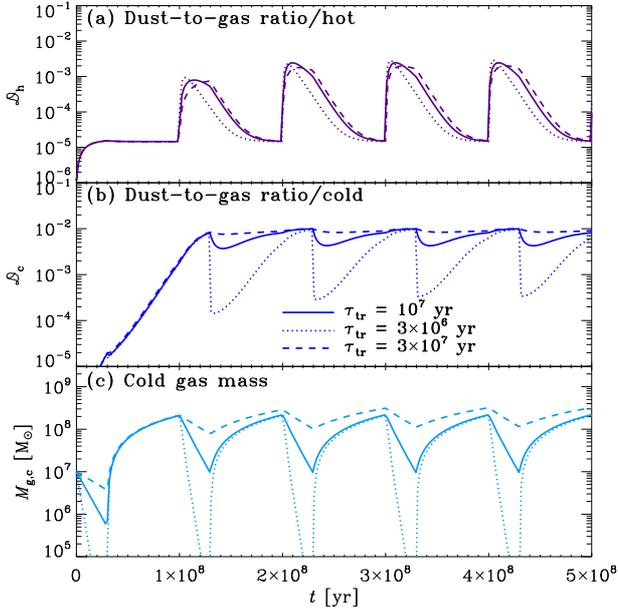}
\caption{Same as Fig.\ \ref{fig:Min} but for various
mass transport times ($\tau_\mathrm{tr}$).
The solid, dotted, and dashed lines show the results for
$\tau_\mathrm{tr}=10^7$ (fiducial), $3\times 10^6$, and
$3\times 10^7~M_\odot$,
respectively, in all the panels.
\label{fig:tautr}}
\end{figure}

The mass transport time-scale of AGN feedback, $\tau_\mathrm{tr}$,
regulates the mass loss rate of the cold gas when AGN feedback
is on (equation \ref{eq:Mret}). In Fig.\ \ref{fig:tautr}, we
show the time evolution of the quantities of interest for
$\tau_\mathrm{tr}=10^7$ (fiducial), $3\times 10^6$, and
$3\times 10^7$ yr.

The increase and decrease of the dust-to-gas ratio in the hot
gas is sharper if $\tau_\mathrm{tr}$ is shorter just because the
effect of mass transport from the cold to hot gas occurs on a
shorter time-scale. The amplitudes of all the quantities shown
in Fig.\ \ref{fig:tautr} are larger for shorter $\tau_\mathrm{tr}$
because larger mass is exchanged between the cold and hot
phases. In particular, the amplitude
of the cold gas mass is sensitive to $\tau_\mathrm{tr}$.
Accordingly, the dust-to-gas ratio in the cold gas drops more
for shorter $\tau_\mathrm{tr}$ because
the effect of dilution by the inflow is more significant.
In contrast, the dust-to-gas ratio in the hot gas
drops to the value insensitive to $\tau_\mathrm{tr}$ when the
AGN feedback is off (i.e., when there is no supply of dust from the
cold gas) because of sputtering. The level of the lowest
$\mathcal{D}_\mathrm{h}$ is determined by the balance between
dust formation by AGB stars and dust destruction by sputtering.

\subsection{Duration of AGN feedback}\label{subsec:fret}

\begin{figure}
\includegraphics[width=0.45\textwidth]{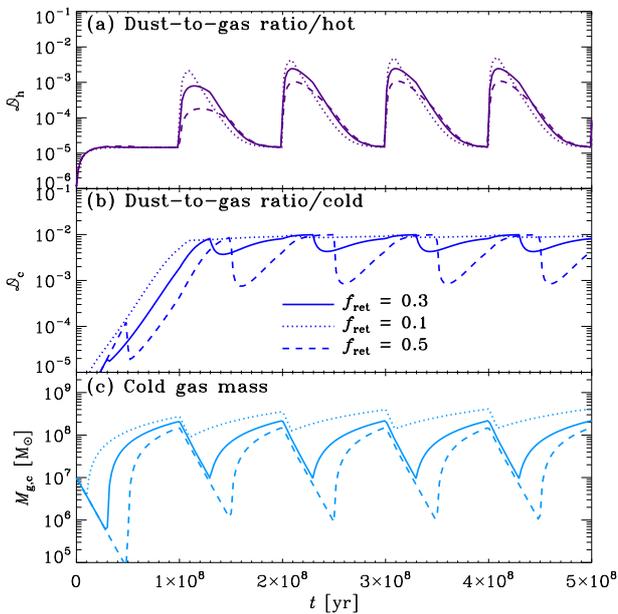}
\caption{Same as Fig.\ \ref{fig:Min} but for various
ratios of the AGN feedback duration to the AGN feedback
duty cycle time ($f_\mathrm{ret}$).
The solid, dotted, and dashed lines show the results for
$f_\mathrm{ret}=0.3$ (fiducial), 0.1, and 0.5,
respectively, in all the panels.
\label{fig:fret}}
\end{figure}

The duration of AGN feedback relative to the AGN duty cycle is
parametrized by $f_\mathrm{ret}$, so that an episode of AGN feedback
lasts for $f_\mathrm{ret}\tau_\mathrm{AGN}$. In Fig.\ \ref{fig:fret},
we show the time evolution of the quantities of interest
for $f_\mathrm{ret}=0.3$ (fiducial), 0.1, and 0.5.

Comparing Figs.\ \ref{fig:tautr} and \ref{fig:fret}, we find
that the duration of AGN feedback has an effect similar to
the mass transport time in terms of the amplitudes:
a larger $f_\mathrm{ret}$ makes the amplitudes of
the cold gas mass and the dust-to-gas ratio in the cold gas
larger. The cold gas naturally decreases more if AGN feedback
lasts longer, and the subsequent dilution of dust abundance
by inflow becomes more significant. The dust-to-gas ratio
in the hot gas has a larger amplitude for a smaller $f_\mathrm{ret}$
since the cold gas mass is larger (i.e., the dust mass supplied from
the cold gas is larger).

\subsection{Dust growth time-scale}

\begin{figure}
\includegraphics[width=0.45\textwidth]{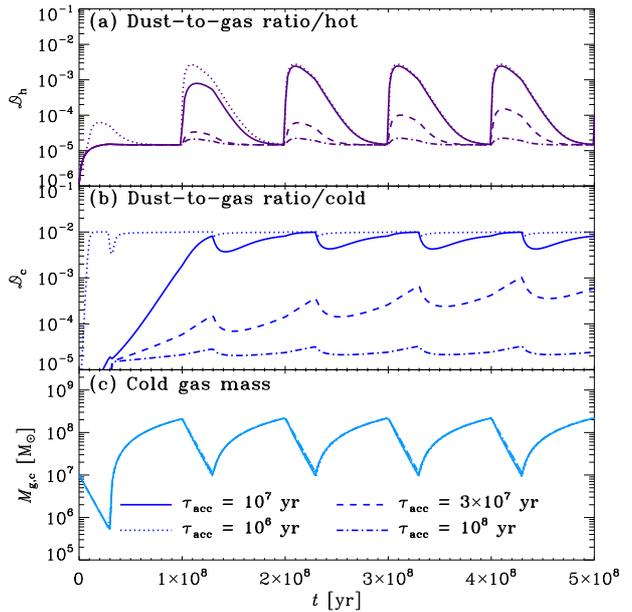}
\caption{Same as Fig.\ \ref{fig:Min} but for various
time-scales of dust growth by accretion ($\tau_\mathrm{acc}$).
The solid, dotted, dashed, and dot-dashed lines show the results for
$\tau_\mathrm{acc}=10^7$ (fiducial), $10^6$, $3\times 10^7$ and $10^8$ yr,
respectively, in all the panels.
\label{fig:tauacc}}
\end{figure}

The accretion time-scale $\tau_\mathrm{acc}$ regulates the
dust growth in the cold gas.
In Fig.\ \ref{fig:tauacc},
we show the time evolution of the quantities of interest
for $\tau_\mathrm{acc}=10^7$ (fiducial), $10^6$,
$3\times 10^7$, and $10^8$ yr.

Naturally, the cold gas mass, which is not related to dust,
is not affected by $\tau_\mathrm{acc}$. As expected, the
dust-to-gas ratio in the cold gas becomes higher for
shorter $\tau_\mathrm{acc}$.
Moreover, if $\tau_\mathrm{acc}$ is as long as
$\tau_\mathrm{AGN}$, the overall level of dust-to-gas 
ratio is suppressed because dust does not grow sufficiently
within an AGN feedback cycle.
The dust-to-gas ratio in the hot gas is also affected,
since the major source of dust in the hot gas is the
dust grown in the cold gas.
We observe a secular increase of $\mathcal{D}_\mathrm{c}$
for $\tau_\mathrm{acc}=3\times 10^7$ yr. This indicates that,
if $\tau_\mathrm{acc}$ is smaller than $\tau_\mathrm{AGN}$,
the dust-to-gas ratio in the cold gas tends to increase unless
it reaches the saturation limit ($\mathcal{D}_\mathrm{sat}$). Therefore,
we observe a bifurcation for the value of $\mathcal{D}_\mathrm{c}$:
if $\tau_\mathrm{acc}<\tau_\mathrm{AGN}$, the dust-to-gas ratio
increases while if $\tau_\mathrm{acc}>\tau_\mathrm{AGN}$, the
dust-to-gas ratio is suppressed and
$\mathcal{D}_\mathrm{c}\sim\mathcal{D}_\mathrm{h}$. In the
latter case, dust growth has little influence on the dust budget
and the dust-to-gas ratio is simply determined by the balance
between dust supply from AGB stars and dust destruction in the
hot gas.

\subsection{Dust destruction time-scale}

\begin{figure}
\includegraphics[width=0.45\textwidth]{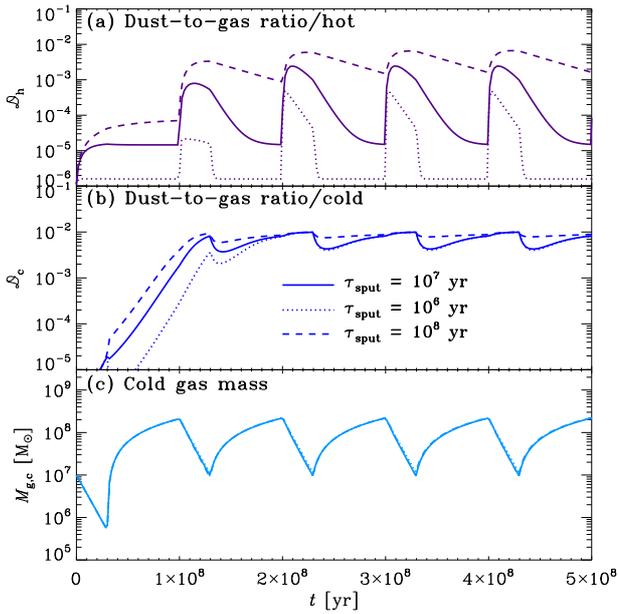}
\caption{Same as Fig.\ \ref{fig:Min} but for various
time-scales of dust destruction by sputtering ($\tau_\mathrm{sput}$).
The solid, dotted, and dashed lines show the results for
$\tau_\mathrm{sput}=10^7$ (fiducial), $10^6$, and $10^8$~yr,
respectively, in all the panels.
\label{fig:tausput}}
\end{figure}

Dust in the hot gas is destroyed by sputtering. The destruction
time-scale is regulated by $\tau_\mathrm{sput}$.
In Fig.\ \ref{fig:tausput}, we show the time evolution of the
relevant quantities for $\tau_\mathrm{sput}=10^7$ (fiducial),
$10^6$, and $10^8$ yr.

Naturally, the cold gas mass, which is not related to dust,
is not affected by $\tau_\mathrm{sput}$. As expected, the dust-to-gas
ratio in the hot gas is largely affected by the sputtering time-scale,
with longer $\tau_\mathrm{sput}$ predicting higher $\mathcal{D}_\mathrm{h}$.
The dust-to-gas ratio in the hot gas is roughly proportional to
$\tau_\mathrm{sput}$ (i.e., inversely proportional to the dust destruction
efficiency).
The dust-to-gas ratio in the cold gas is also kept high for
$\tau_\mathrm{sput}=10^8$ yr: if $\mathcal{D}_\mathrm{h}$
remains as high as $\mathcal{D}_\mathrm{c}$, 
the dust-to-gas ratio
in the cold gas is kept high because the dust-to-gas ratio in the
infalling gas is high (i.e., the dust abundance in the cold gas
is not diluted by the infall). This effect is significant only if
$\tau_\mathrm{sput}\gtrsim\tau_\mathrm{AGN}$.

\subsection{Stellar mass}

\begin{figure}
\includegraphics[width=0.45\textwidth]{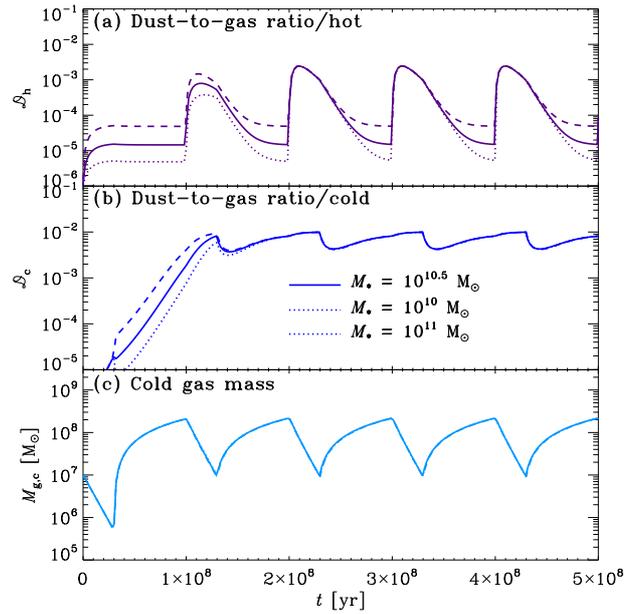}
\caption{Same as Fig.\ \ref{fig:Min} but for various
stellar masses ($M_*$).
The solid, dotted, and dashed lines show the results for
$M_*=10^{10.5}$ (fiducial), $10^{10}$, and $10^{11}~M_\odot$,
respectively, in all the panels.
\label{fig:Mstar}}
\end{figure}

The total stellar mass affects the dust evolution through
dust production by AGB stars. In Fig.\ \ref{fig:Mstar},
we show the time evolution of the relevant quantities
for $M_*=10^{10.5}$ (fiducial), $10^{10}$, and $10^{11}~M_\odot$.

The stellar mass does not affect the cold gas mass in our model
since it only contributes to the dust production.
Because the dust supply from AGB stars to the hot gas is
proportional to the stellar mass, the lowest level of dust-to-gas
ratio which is determined by the balance between dust
supply and dust destruction is proportional to the stellar mass.
The maximum level of the dust-to-gas ratio in the hot gas
is insensitive to the stellar mass, because it is determined
by the dust injection from the cold gas. The dust-to-gas ratio
in the cold gas is not sensitive to the stellar mass as long
as dust growth is efficient enough.

\section{Discussion}\label{sec:discussion}

In this section, we mainly compare the theoretical calculations with
observational data. Here we utilize the observational data for
dust and gas in elliptical galaxies. Because of the
general deficiency of gas and dust in elliptical galaxies,
data is limited, while
the stellar mass is relatively easy to be determined in these
systems using optical photometric observations.
Thus, we also adopt the stellar mass and utilize it
(mainly for normalization).

\subsection{Sample}\label{subsec:sample}

One of the newest systematic dust observations in
nearby elliptical galaxies is carried out by \textit{Herschel}.
We adopt the sample from \citet{Smith:2012aa} and
\citet{di-Serego-Alighieri:2013aa}. The sample galaxies adopted are
listed in Table \ref{tab:sample}. We selected galaxies with
a morphological type of elliptical galaxies (E) and excluded
objects with stellar mass with $<10^{9.5}~M_\odot$.
NGC 4374 is common for both samples. The stellar mass is taken
from \citet{Cortese:2012aa}. The gas masses are traced by H \textsc{i}
and H$_2$ (CO); therefore, the observed gas masses are compared
with the cold gas mass in our models.
We only compare the observed dust mass with the dust mass in the
cold gas in our models, because the observational sensitivity
to the diffuse component in the hot gas depends on the
spatial extension, which is not known. In any case, since
the dust mass in the cold gas is much larger than that in the
hot gas in most of the time, adding the dust mass in the hot gas
does not significantly change the comparison below.

\begin{table*}
\centering
\begin{minipage}{180mm}
\caption{Elliptical galaxy sample adopted from \citet{Smith:2012aa}
and \citet{di-Serego-Alighieri:2013aa}.}
\label{tab:sample}
\begin{center}
\begin{tabular}{@{}llcccccc} \hline
Name & Other name & $\log M_*$ &  $\log M_\mathrm{dust}$ &
$\log M_\mathrm{HI}$ & $\log M_\mathrm{H_2}$ & $\log\mathcal{D}$ & Ref.$^a$\\
 & & ($M_\odot$) & ($M_\odot$) & ($M_\odot$) &
($M_\odot$) & \\ \hline
VCC 763 & NGC 4374, M84 & 11.18 & 5.05$^b$ & 8.96 & $<$7.23 &
$-$3.91$^c$ & 1, 2, 3\\
HRS 150 & NGC 4406, M86 & 11.22 & 6.63 & 7.95 & $<$7.4 &
$-$1.43$^c$ & 1, 3\\
HRS 186 & NGC 4494 & 10.88 & 5.08 & 8.26 & $<$7.35 &
$-$3.23$^c$ & 1, 3\\
HRS 241 & NGC 4636 & 10.98 & 5.06 & 9.0 & $<$7.02 &
$-$3.94$^c$ & 1, 3 \\
VCC 345 & NGC 4261 & 11.32 & 5.81 & $<$8.45 & $<$7.70 &
$>-$2.71$^d$ & 2, 3\\
VCC 881C & NGC 4406 & 11.22 & 5.47 & 7.95 & --- & $-$2.48$^e$ & 2, 3\\
VCC 1226 & NGC 4472, M49 & 11.39 & 5.49 & $<$7.90 & $<$7.26 &
$>-$2.50$^d$ & 2, 3\\
VCC 1619 & NGC 4550 & 10.03 & 5.41 & $<$7.90 & 7.20 &
$>-$2.57$^d$ & 2, 3\\
\hline
HRS 3 & NGC 3226 & 10.21 & 5.96 & --- & $<7.13$ & --- & 1, 3\\
HRS 43 & NGC 3608 & 10.27 & $<4.88$ & --- & $<7.28$ & --- & 1, 3\\
HRS 49 & NGC 3640 & 10.70 & $<5.20$ & --- & $<7.25$ & --- & 1, 3\\
HRS 125 & NGC 4339 & 10.30 & $<5.17$ & $<7.84$ & $<7.2$ & --- & 1, 3\\
HRS 135 & NGC 4365 & 11.48 & $<6.17$ & $<8.18$ & $<7.62$ & --- & 1, 3\\
HRS 179 & NGC 4473 & 10.71 & $<5.19$ & $<7.92$ & $<7.16$ & --- & 1, 3\\
HRS 181 & NGC 4478 & 10.09 & $<4.85$ & --- & $<7.31$ & --- & 1, 3\\
HRS 211 & NGC 4552, M89 & 10.80 & $<5.67$ & $<7.92$ & $<7.36$ & --- & 1, 3\\
HRS 214 & NGC 4564 & 10.25 & $<5.30$ & $<7.79$ & $<7.33$ & --- & 1, 3\\
HRS 218 & NGC 4570 & 10.48 & $<5.67$ & $<7.31$ & $<7.47$ & --- & 1, 3\\
HRS 236 & NGC 4621, M59 & 10.99 & $<5.76$ & $<7.92$ & $<7.24$ & --- & 1, 3\\
HRS 245 & NGC 4649, M60 & 11.34 & $<5.40$ & $<7.92$ & $<7.59$ & --- & 1, 3\\
HRS 248 & NGC 4660 & 10.05 & $<4.85$ & $<7.92$ & $<7.30$ & --- & 1, 3\\
HRS 258 & NGC 4697 & 11.10 & 5.46 & --- & $<7.25$ & --- & 1, 3\\
HRS 312 & NGC 5576 & 10.60 & $<5.51$ & --- & $<7.46$ & --- & 1, 3\\
HRS 316 & NGC 5638 & 10.52 & $<5.23$ & --- & $<7.54$ & --- & 1, 3\\
VCC 1316 & NGC 4486, M87 & 11.24 & 5.34 & --- & $<7.18$ & --- & 2, 3\\
VCC 1327 & NGC 4486A & 10.05 & 4.92 & --- & --- & --- & 2, 3\\
\hline
\end{tabular}
\end{center}
$^a$References: 1) \citet{Smith:2012aa};
2) \citet{di-Serego-Alighieri:2013aa}; 3) \citet{Cortese:2012aa}.\\
$^b$Adopted from Ref.\ 1. Ref.\ 2 gives 5.30 with the same mass
absorption coefficient.\\
$^c$The upper limit of H$_2$ mass is used, but this does not
affect the estimated dust-to-gas ratio since the gas mass is
dominated by H\,\textsc{i} gas mass.\\
$^d$We use the upper limits of gas mass to derive the lower
limits of dust-to-gas ratio.\\
$^e$We neglect the H$_2$ mass for the gas mass.
\end{minipage}
\end{table*}

\subsection{Cold gas mass}\label{subsec:cold}

The variation of the cold gas mass in our models
in Section~\ref{sec:result} is broadly in the range of
the observed gas masses shown in Table \ref{tab:sample}
($\sim 10^9~M_\odot$ down to $\sim 10^7~M_\odot$ or less).
Therefore, we confirm that the choices of parameter values in
Section \ref{sec:model} are reasonable as far as the cold gas mass
is concerned. The largest cold gas mass is realized if we
adopt the largest inflow rate ($\dot{M}_\mathrm{in,0}=10~M_\odot$ yr$^{-1}$;
Fig.\ \ref{fig:Min}). Therefore, we confirm that the maximum inflow rate
by the accretion of the cooled gas is $\lesssim 10~M_\odot~\mathrm{yr}^{-1}$.
The amplitude of the cold gas mass is, on the other hand, determined
by the time-scale of gas transport in AGN feedback (Fig.~\ref{fig:tautr})
and the duration of AGN feedback (Fig.~\ref{fig:fret}).
If the variation
of the gas mass in the sample is interpreted as due to time evolution,
$\tau_\mathrm{tr}\gtrsim 3\times 10^7$ yr or $f_\mathrm{ret}\lesssim 0.1$ is
not favored because the variation of the cold gas is smaller
than an order of magnitude.
The non-detection of H\,\textsc{i} and H$_2$ gas for a large part of
the sample galaxies is consistent with even the most extreme case
of $\tau_\mathrm{tr}=3\times 10^6$~yr or $f_\mathrm{ret}=0.5$. This means that
it is difficult to constrain the lower bound for $\tau_\mathrm{tr}$
and the upper bound for $f_\mathrm{ret}$. 

\subsection{Dust mass}

The dust mass in the cold gas varies between $\sim 10^5$ and $\sim 10^6~M_\odot$
in the fiducial model (Fig.\ \ref{fig:std}). The maximum of the dust mass is achieved
when the cold gas mass becomes the largest (i.e., just before AGN feedback).
In this phase, the dust-to-gas ratio is almost saturated if $\tau_\mathrm{acc}$ is
significantly shorter than $\tau_\mathrm{AGN}$. If we adopt the maximum
value of the dust-to-gas ratio ($\mathcal{D}_\mathrm{sat}=0.01$), we
obtain the maximum dust mass
in the case of the largest cold gas mass ($10^9~M_\odot$; see above) as
$\sim 10^7~M_\odot$. The observed dust mass is indeed smaller than this
value. Because the logarithmic variation of the dust-to-gas ratio is smaller than that
of the cold gas mass, the time evolution of the dust mass is dominated by the
change of the cold gas mass in our models.

\subsection{Dust-to-stellar mass ratio}\label{subsec:ds_ratio}

To cancel out the size effects of elliptical galaxies, it is useful to
normalize the dust mass to an indicator of total mass scale.
Here we use the stellar mass for the normalization; that is,
we use the dust-to-stellar mass ratio for the indicator of the richness
of dust. As an indicator of AGN feedback, we also show the
ratio of the cold gas mass to the stellar mass, and examine
the relation between $M_\mathrm{d,c}/M_*$ and $M_\mathrm{g,c}/M_*$.
For the data points, we exclude the objects without any constraint on
the H \textsc{i} mass,
since it is impossible to constrain the gas mass in such a case.
If H \textsc{i} is detected, we sum the H \textsc{i} mass and the H$_2$ mass
to obtain the total cold gas mass (if H$_2$ is not detected, we add the
upper limit of H$_2$ gas mass to the H \textsc{i} gas mass).

\begin{figure*}
\begin{center}
\includegraphics[width=0.45\textwidth]{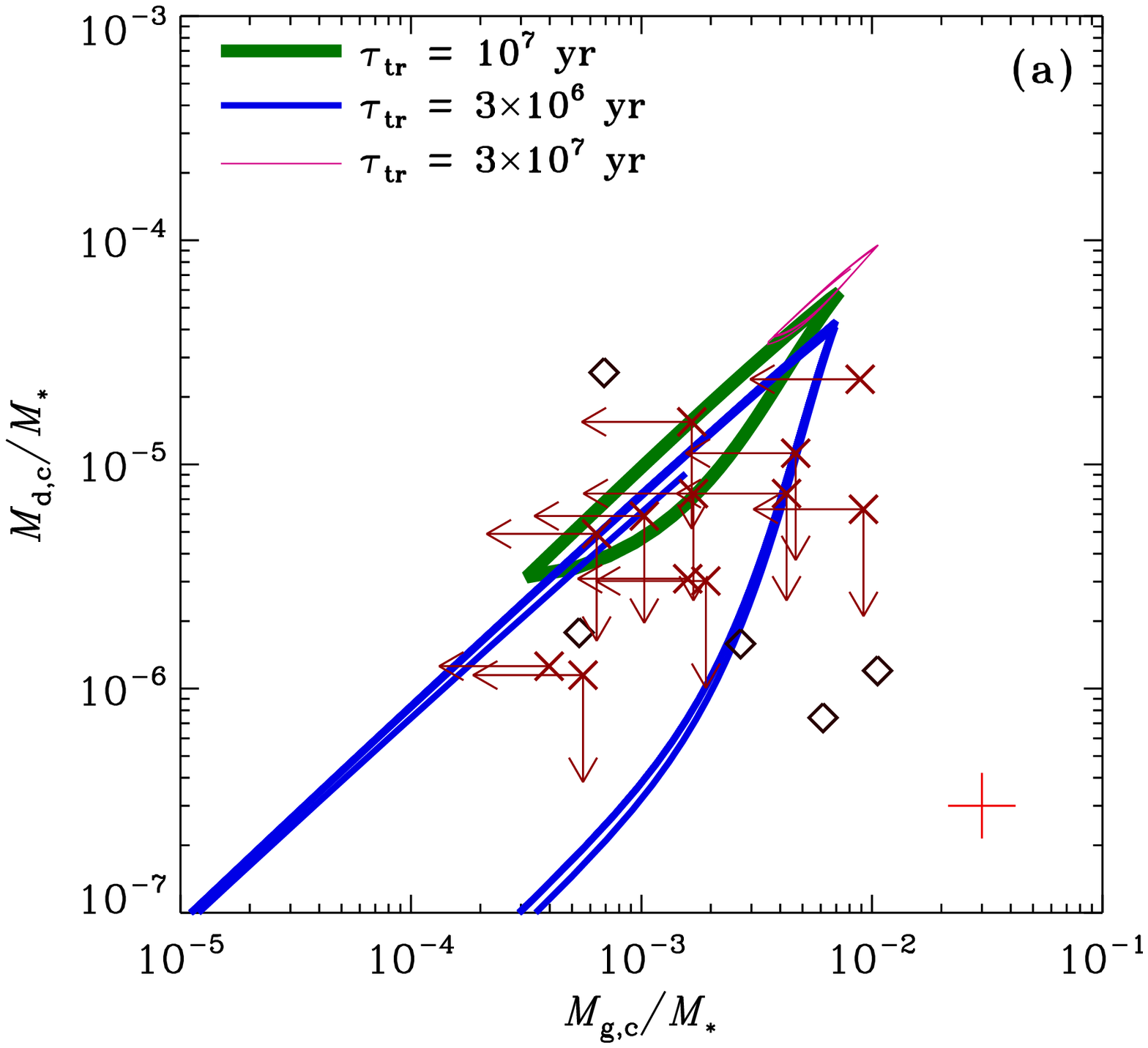}
\includegraphics[width=0.45\textwidth]{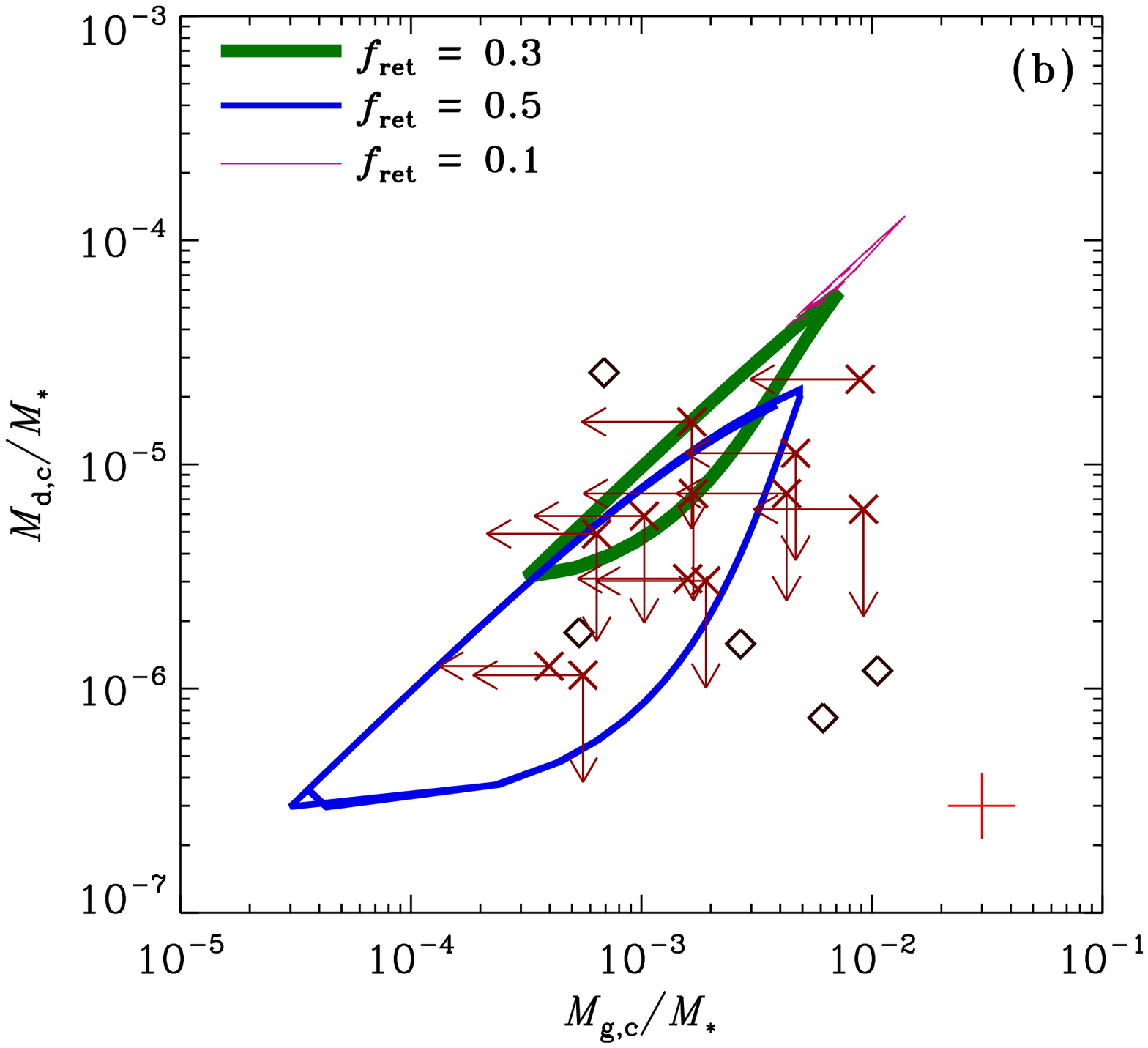}\\
\includegraphics[width=0.45\textwidth]{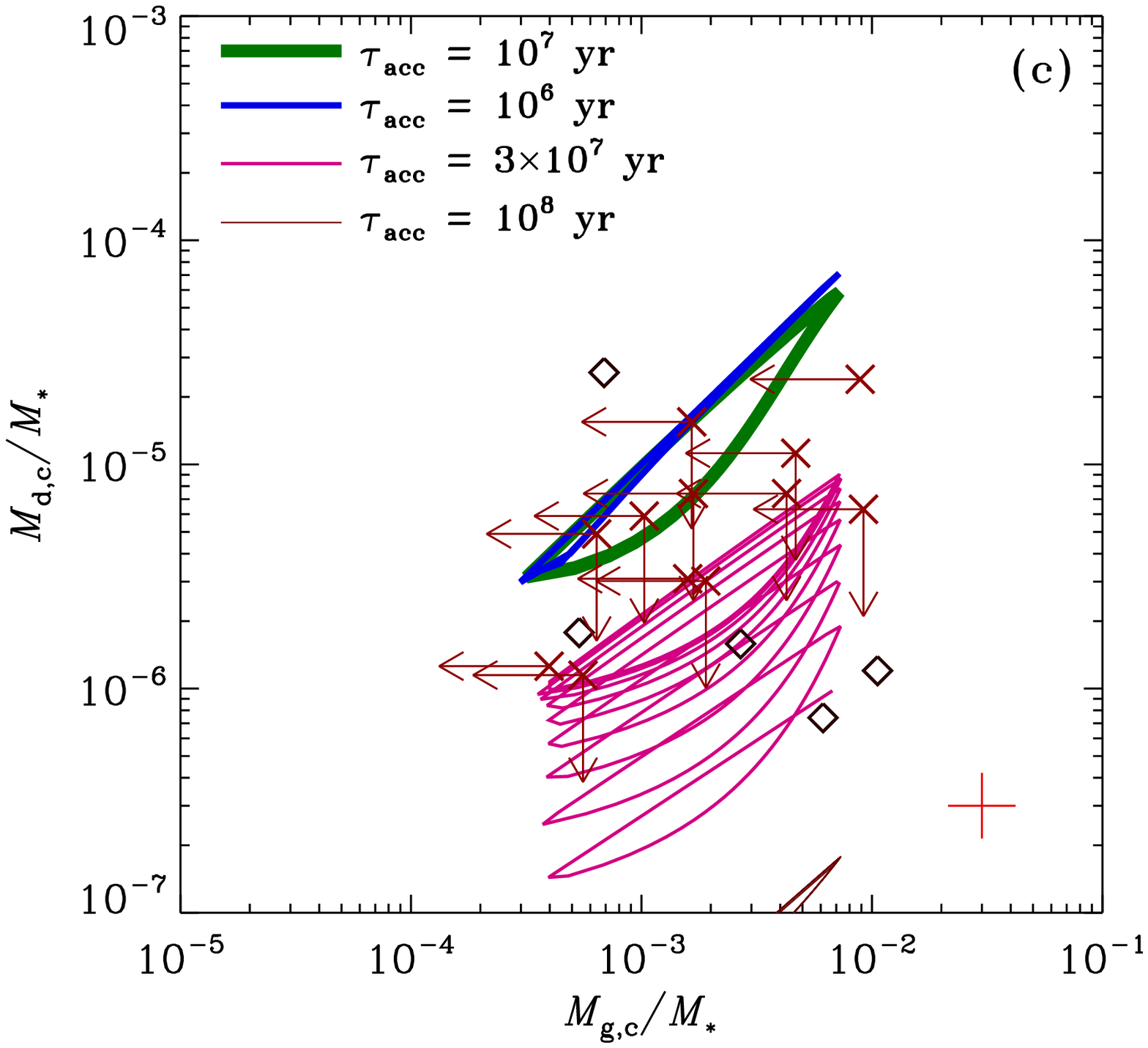}
\includegraphics[width=0.45\textwidth]{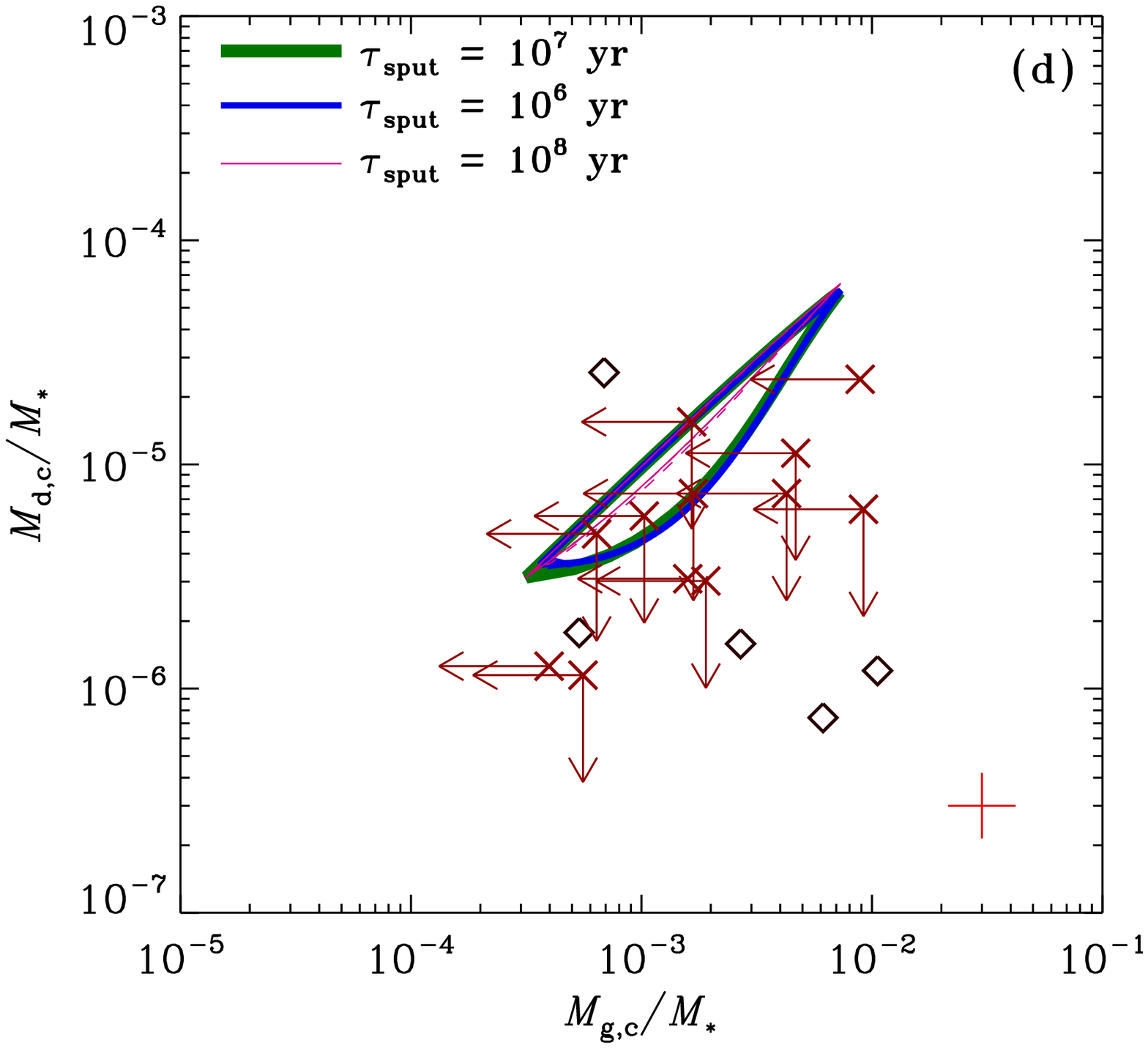}
\end{center}
\caption{Relation between dust-to-stellar mass ratio
($M_\mathrm{d,c}/M_*$) and gas-to-stellar mass ratio
($M_\mathrm{g,c}/M_*$).
(a) Dependence on $\tau_\mathrm{tr}$.
The thick (green), medium (blue) and thin (red) lines represent the trajectory
for $\tau_\mathrm{tr}=10^7$ (fiducial), $3\times 10^6$, and
$3\times 10^7$ yr, respectively.
(b) Dependence on $f_\mathrm{ret}$.
The thick (green), medium (blue), and thin (red) lines show the trajectory
for $f_\mathrm{ret}=0.3$ (fiducial), 0.5, and 0.1, respectively.
(c) Dependence on $\tau_\mathrm{acc}$.
The thick (green), medium thick (blue), medium thin (red), and thin (brown)
lines represent the trajectory
for $\tau_\mathrm{acc}=10^7$ (fiducial) $10^6$, $3\times 10^7$, and
$10^8$ yr, respectively.
(d) Dependence on $\tau_\mathrm{sput}$.
The thick (green), medium (blue), and thin (red) lines show the trajectory for
$\tau_\mathrm{sput}=10^7$ (fiducial), $10^6$, and $10^8$ yr,
respectively. These three lines are almost overlapped.
We also show observational data points in all
the panels listed in Table \ref{tab:sample}: The diamonds show the
galaxies for which both dust mass and gas mass are detected, while
the crosses with arrows present the data points for which either dust
mass or gas mass is not detected (only an upper limit is obtained).
The cross on the right lower corner shows the typical error for the
observational data points.
\label{fig:cycle_star}}
\end{figure*}

In Fig.~\ref{fig:cycle_star}, we show the relation between
$M_\mathrm{d,c}/M_*$ and $M_\mathrm{g,c}/M_*$.
We only show the dependence on the parameters which affect
the relative abundance of dust to gas (i.e., dust-to-gas ratio).
Those parameters are $\tau_\mathrm{tr}$, $f_\mathrm{ret}$,
$\tau_\mathrm{acc}$ and $\tau_\mathrm{sput}$.
The same ranges for the parameters as above
(i.e., the ranges in Table \ref{tab:param}) are investigated.
We only plot the relation at $t>2\times 10^8$~yr, when the
trajectory almost converges to a limit cycle (by ``forgetting the
initial condition'') except for the case with
$\tau_\mathrm{acc}=3\times 10^7$~yr in Panel (c). In this case,
there is a secular drift toward high dust abundance as shown in
Fig.\ \ref{fig:tauacc}.

Overall, we observe in Fig.\ \ref{fig:cycle_star} that the
variation is dominated by the change of the gas mass because
the variation is mostly in the diagonal direction on the diagram.
If the gas mass changes, both $M_\mathrm{d,c}/M_*$ and $M_\mathrm{g,c}/M_*$
moves on the diagonal line under a constant $\mathcal{D}_\mathrm{c}$.
In other words, the variation of the dust-to-gas ratio can be
seen in off-diagonal behavior. Indeed, such off-diagonal behavior
is seen clearly for a short $\tau_\mathrm{tr}$ in Fig.\ \ref{fig:cycle_star}a and
a large $f_\mathrm{ret}$ in Fig.\ \ref{fig:cycle_star}b.
In both cases, the cold gas mass is lost drastically in the epoch of
AGN feedback, so that the dilution of dust abundance by the inflow of dust-poor
gas has a large impact on the dust-to-gas ratio in the cold gas
(Sections \ref{subsec:Min} and \ref{subsec:fret}).

We observe in Fig.\ \ref{fig:cycle_star} that a single model does
not cover an area wide enough to explain all the data points.
In particular, the data points at high $M_\mathrm{g,c}/M_*$ and
low $M_\mathrm{d,c}/M_*$ are difficult to be reproduced.
As shown above, the dilution effect appears just after the epoch of
AGN feedback, when the cold gas mass is low. Therefore,
it is difficult to reproduce the low dust content at the high
gas mass. In our scenario, the only way to reproduce such a
dust-poor and gas-rich data point is to change the efficiency of
accretion as shown in Fig.\ \ref{fig:cycle_star}c.
The same effect is also realized by lowering $\mathcal{D}_\mathrm{sat}$
(equation \ref{eq:taugrow}).
Therefore, the large dispersion of dust mass at high
gas mass can be interpreted as various dust growth efficiencies.
A variety in dust growth efficiency could be caused by a variety in
dense gas fraction, cold gas density, metallicity, or
grain size distribution (Section \ref{subsec:param}).

Some studies argue that the dust-to-stellar mass ratio changes
as the system is enriched with dust
\citep[e.g.,][]{Remy-Ruyer:2015aa}.
Dust enrichment is indeed the most important factor of
changing the dust-to-stellar mass ratio in gas-rich star-forming galaxies,
since dust production by stars and dust growth in the cold dense
ISM are actively occurring. In this paper, we have shown that,
in elliptical galaxies, the dust-to-stellar mass ratio is still
varied by dust growth, although it is driven by a cycle of AGN feedback,
not by chemical enrichment. Therefore, our results imply that,
if AGN feedback affects the galaxy evolution, we also need to
consider the variation of dust abundance by AGN feedback.

\subsection{Dust-to-gas ratio}

As shown above, the behavior on the
$M_\mathrm{d,c}/M_*$--$M_\mathrm{g,c}/M_*$ diagram is dominated
by the overall oscillation of the cold gas mass. To examine the
effect of dust abundance variation more clearly, it is useful to
consider a quantity normalized to the gas mass. In this way, we
can cancel the effect of oscillatory behavior of the cold gas mass,
and concentrate on the variation of dust abundance. We still use
the gas-to-stellar mass ratio, $M_\mathrm{g,c}/M_*$, for the
horizontal axis
to show the cycle of AGN feedback. Thus, we consider the
$\mathcal{D}_\mathrm{c}$--$M_\mathrm{g,c}/M_*$ diagram here.
The same observational sample as in
Section~\ref{subsec:ds_ratio} is chosen for comparison but we excluded
galaxies without detection of dust emission.

\begin{figure*}
\begin{center}
\includegraphics[width=0.45\textwidth]{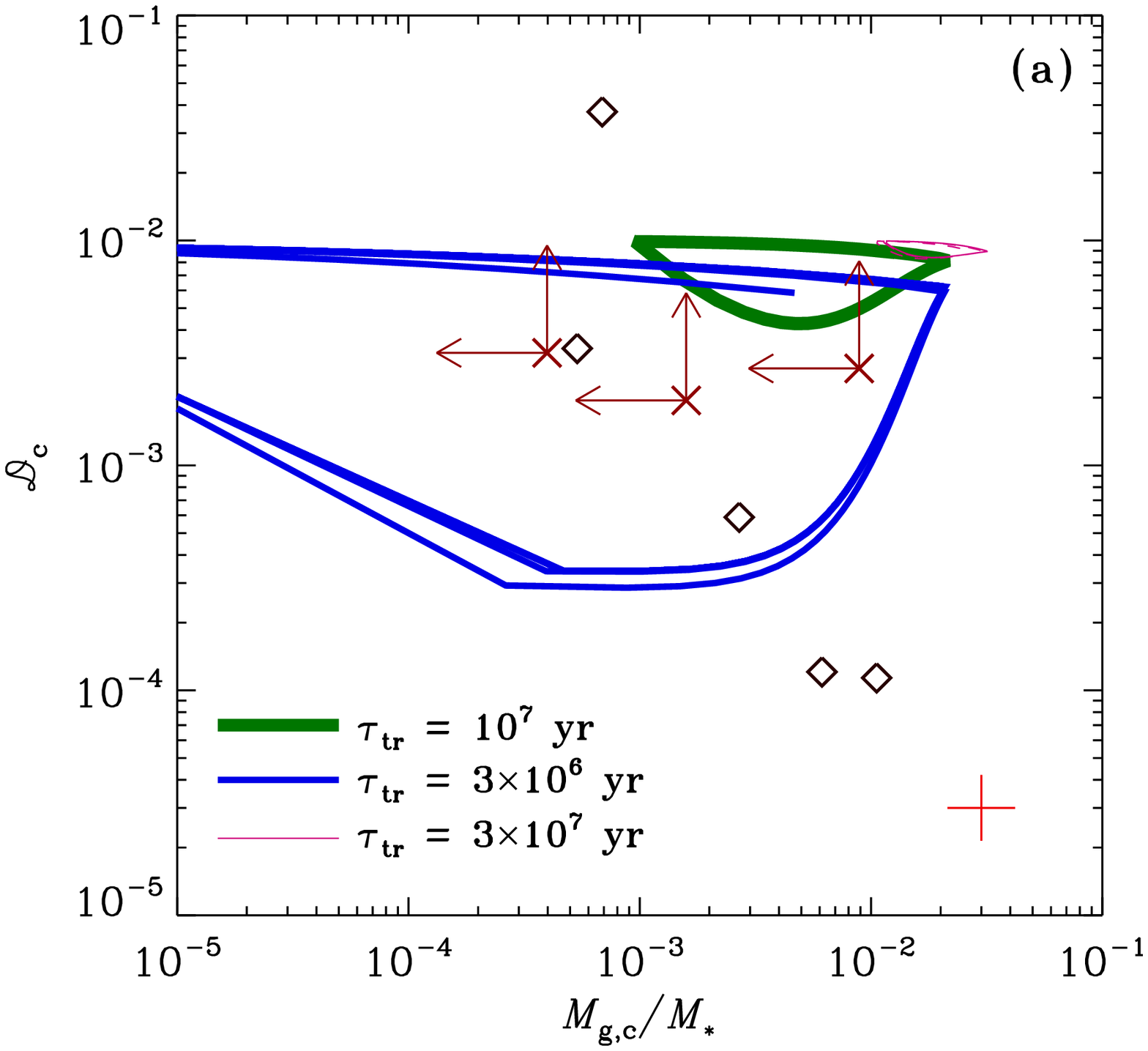}
\includegraphics[width=0.45\textwidth]{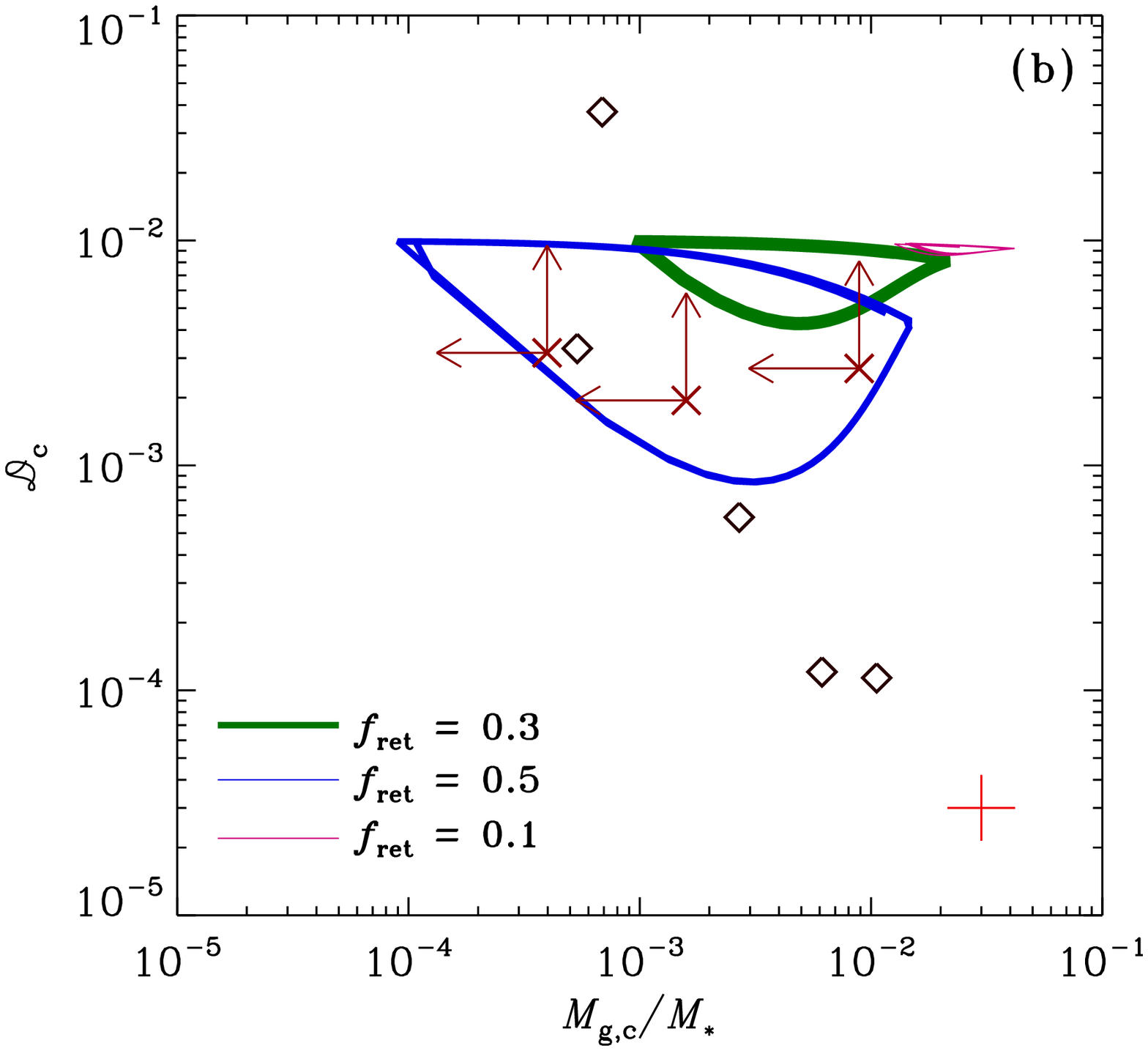}\\
\includegraphics[width=0.45\textwidth]{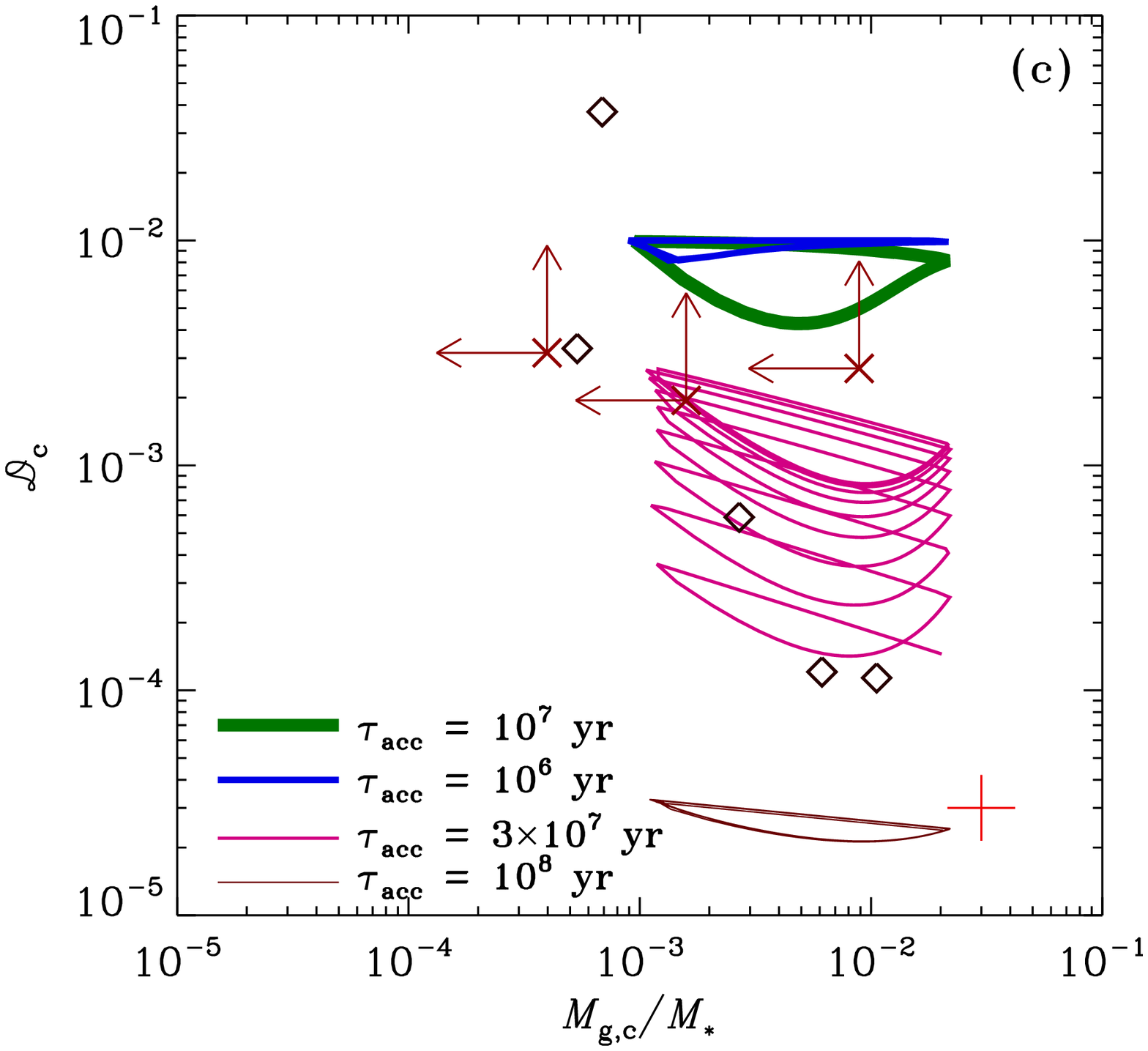}
\includegraphics[width=0.45\textwidth]{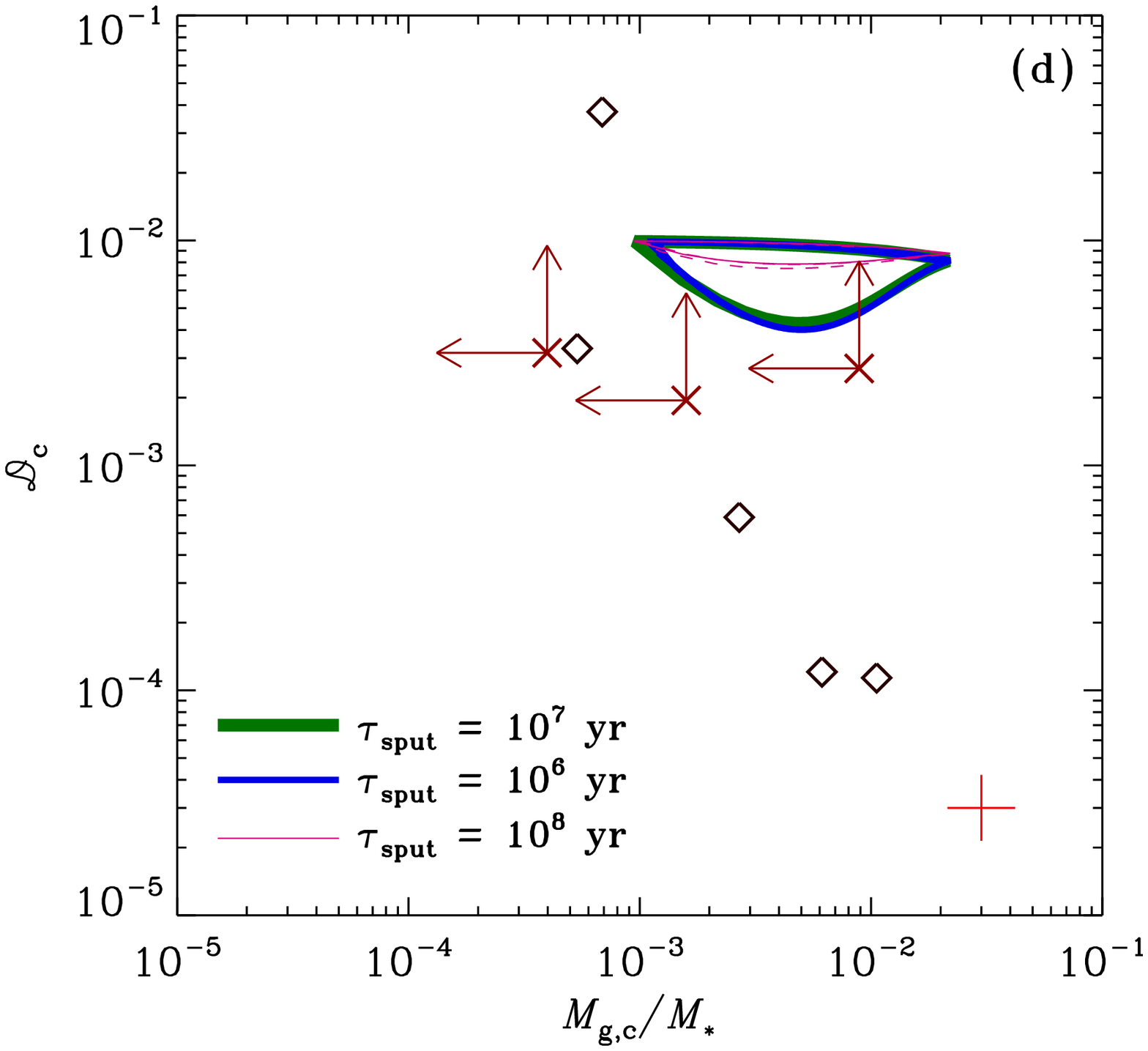}
\end{center}
\caption{Same as Fig.\ \ref{fig:cycle_star} but showing the
relation between dust-to-gas ratio
($\mathcal{D}_\mathrm{c}=M_\mathrm{d,c}/M_\mathrm{g,c}$) and
gas-to-stellar mass ratio ($M_\mathrm{g,c}/M_*$).
We also show observational data points listed in Table \ref{tab:sample}
(but only the objects with dust detection are shown):
The diamonds show the
galaxies for which gas is detected, while
the crosses with arrows present the data points without gas
detection (only an upper limit is obtained for the gas mass).
The cross on the right lower corner shows the typical error.
\label{fig:cycle_gas}}
\end{figure*}

In Fig.\ \ref{fig:cycle_gas}, we show the relation between
$\mathcal{D}_\mathrm{c}$ and $M_\mathrm{g,c}/M_*$.
We only plot the relation at $t>2\times 10^8$ yr as also done
in Fig.\ \ref{fig:cycle_star}. We examine the variations of
the same parameters as in
Section \ref{subsec:ds_ratio}. The horizontal axis shows the
``phase'' of AGN feedback
with high and low phases of $M_\mathrm{g,c}/M_*$ showing
the timings when AGN feedback is turned on and off,
respectively. The upper limit of
$\mathcal{D}_\mathrm{c}$ is
determined by the saturation of dust growth
($\mathcal{D}_\mathrm{sat}$). After AGN
feedback, $\mathcal{D}_\mathrm{c}$ becomes low by
the dilution as a result of the inflow of dust-poor gas.
As the cold gas mass increases, dust growth raises
the dust-to-gas ratio.

As concluded in Section \ref{subsec:ds_ratio}, it is difficult to
reproduce the dust-poor objects at high gas mass, and the only
way to reproduce them in our model is to assume inefficient dust
growth. However, it should be emphasized that our model is capable
of reproducing an order of magnitude variation in dust-to-gas ratio
if AGN feedback efficiently evaporates the cold gas into the hot gas
($\tau_\mathrm{tr}=3\times 10^6$ yr in Fig.\ \ref{fig:cycle_gas}a
and $f_\mathrm{ret}=0.5$ in Fig.\ \ref{fig:cycle_gas}b).

\section{Conclusion}\label{sec:conclusion}

In this paper, we modeled the evolution of dust in elliptical galaxies
and investigated the dust formation and evolution in the cold
and hot gas components. We also considered the mass exchange
between the two gas components by including AGN feedback, which
heats the cold gas and converts it to hot gas, and also
by considering cooling of the hot gas, which is accreted on to the
cold gas. We considered a new dust formation path through dust growth by
the accretion of gas-phase metals in addition to dust formation by AGB
stars. Thus, the
injection of the cold gas into the hot gas by AGN feedback
acts as a dust supplying mechanism to the hot gas.
We also included sputtering in the hot gas following previous
papers. Since sputtering is efficient enough to make the dust-to-gas
ratio in the hot gas much lower than that in the cold gas,
the mass inflow to the cold gas dilutes the dust abundance there.

We surveyed reasonable ranges of the relevant parameters in
the model and clarified how each process affects the evolution
of dust and gas in elliptical galaxies. (i) The mass inflow rate
into the cold gas naturally affects the cold gas mass, although
it has a minor influence on the dust-to-gas ratio in the cold gas.
Therefore, the role of mass inflow is to characterize the total
mass scale of the cold gas and dust. (ii) The mass transport
time-scale from the cold gas to the hot gas in AGN feedback
affects the amplitudes of the cold gas mass and the dust-to-gas
ratio in the cold gas. If this time-scale is shorter, the cold gas
loses a larger fraction of its mass in AGN feedback. As a result,
the subsequent inflow of dust-poor gas dilutes the
dust abundance in the cold gas more. Therefore, the variation of
dust-to-gas ratio in the cold gas is larger for a shorter
mass transport time-scale.
(iii) The duration of AGN feedback also regulates the
cold gas lost in an AGN feedback cycle. Therefore, its effect
is similar to that of the mass transport time-scale, and
the amplitude of the dust-to-gas ratio in the cold gas
is larger for a longer duration of AGN feedback.
(iv) The dust growth time-scale in the cold gas has the largest
impact on the dust-to-gas ratio there. It affects the
absolute level of dust-to-gas ratio rather than the amplitude
of dust-to-gas ratio. The low dust-to-gas ratios in some relatively gas-rich
elliptical galaxies could be most naturally explained by a large variation
in dust growth time-scale among the sample. Such a variation
can be a consequence of a variety in dense gas fraction,
grain size distribution, metallicity, or gas density.
(v) As long as the sputtering time-scale is significantly
shorter than the AGN feedback cycle, dust in the hot gas is
destroyed efficiently so that the infall from the hot to the cold
gas phase always works as
dilution of the dust abundance in the cold gas. However, if
the sputtering time-scale is comparable or longer than the AGN feedback cycle,
the dilution of the dust abundance in the cold gas does not
occur efficiently (i.e., the dust-to-gas ratio in the hot gas
remains high), so that the dust-to-gas ratio in the cold gas
does not drop after the inflow.
(vi) Because the major source of dust is dust growth in the
cold gas, dust production by AGB stars does not affect the
evolution of the dust-to-gas ratio in the cold gas.

By comparing with observational data of nearby elliptical
galaxies, we showed that the variety in gas mass is
nicely explained by our models and that the variety in the dust-to-gas ratio
can also be reproduced to some extent. However, the full range
of the observed dust-to-gas ratio is
difficult to be reproduced unless we vary the dust growth
time-scale (or dust growth efficiency). In our framework, the low dust-to-gas ratios
in relatively gas-rich objects are only reproduced with
inefficient dust growth (or long dust growth time-scales).
Therefore, we conclude that dust growth can play a central role in
creating the variation in dust-to-gas ratio through the AGN feedback
cycle and through the variation in dust growth efficiency.

\section*{Acknowledgment}

We are grateful to T. Kokusho and H. Kaneda
for their useful comments and discussions, and to the
anonymous referees for their careful reviews of the paper.
HH is supported by the Ministry of Science and Technology
(MoST) grant 102-2119-M-001-006-MY3.
TN is supported in part by the JSPS Grant-in-Aid for
Scientific Research (26400223).



\bibliographystyle{elsarticle-harv} 
\bibliography{hirashita1}

\end{document}